\documentstyle[sprocl,amsbsy,latexsym]{article}
\newcommand{\bra}[1]{\langle #1 \vert}
\newcommand{\ket}[1]{\vert #1 \rangle}
\newcommand{\mod}[1]{\vert {#1}\vert}

\begin{document}
\setlength{\unitlength}{1mm}

\title{Colour Charges and Anti-Screening}

\author{E.\ Bagan}

\address{Grup de F\'{\i}sica Te\`orica and IFAE\\
Edificio Cn, Universitat Aut\`onoma de Barcelona\\ E-08193
Bellaterra\\Spain\\E-mail: bagan@ifae.es}

\author{R.\ Horan, M.\ Lavelle and D.\ McMullan\footnote{Talk presented  by D.\ McMullan.}}

\address{School of Mathematics and Statistics\\
University of Plymouth\\ Drake Circus, Plymouth, Devon PL4
8AA\\U.K.\\E-mail: rhoran/mlavelle/dmcmullan@plymouth.ac.uk}

\maketitle\abstracts{If constituent quarks are to emerge from QCD
they must have well defined colour and be energetically favoured.
After reviewing the general properties of charges in gauge
theories, a method for constructing charges is  presented and
applied to the infra-red structure of the theory and to the
interquark potential. Both of these applications supply a physical
interpretation of the structures found in the construction of
charges. We will see that constituent structures arise in QCD.}

\section*{Asymptotic Dynamics}

The S-matrix is used to describe the scattering of in-particle
states to some final out-particle states. The existence of such an
S-matrix description of particle scattering is  based on the
overlap between the interacting and free theory in the remote past
and future, i.e., in QED with the interaction Hamiltonian
\begin{equation}
H_I(t)=-{\mathrm{e}}\int\! d^3
x\,\overline{\psi}(x)\gamma^\mu\psi(x)A_\mu(x)\,,
\end{equation}
it is assumed that $H_I(t)\to 0$ as $t\to\pm\infty$. This large time limit,
though, fails to be that of the free theory in QED. Instead one finds
that the interaction survives:
\begin{equation}
H_I(t)\to -{\mathrm{e}}\int\! d^3x\,
A^\mu(x)J^{\mathrm{as}}_\mu(x)\,,
\end{equation}
where the non-vanishing asymptotic current has the form\,\cite{kulish:1970}
\begin{equation}
J^{\mathrm{as}}_\mu(x)=\int\!
d^3p\,\rho(p)\frac{p^\mu}{E_p}\delta^{(3)}(\underline{x}-\frac{t
\underline{p}}{E_p})\,.
\end{equation}
What this is telling us is that in such a theory with massless
fields, the associated long-range nature of the interactions
implies that we cannot simply \lq\lq switch off the
coupling\rq\rq. The asymptotic dynamics for the matter  is not
free, and  hence there is no particle description at early or late
times. This is the well-known infra-red problem in QED.

The interpretation of this result that we wish to emphasize,
though,  is that  Gauss' law is not trivial asymptotically: the
matter and gauge fields cannot be viewed as independent. In
particular, the matter fields are not physical as they are not
separately  gauge invariant even at remote times. This raises the
obvious question: if the matter fields are not physical, then what
are the physical charges in this theory, and can they have a
particle description?

\section*{Construction of Charges}

Charges must be gauge invariant to be physical. The generic form
for a charged field in QED is thus given by the product of the
form $h^{-1}(x)\psi(x)$ where, under a gauge transformation,
$\psi\to e^{i{\mathrm{e}}\theta}\psi$, the dressing transforms as
\begin{equation}\label{ginv}
h^{-1}\to h^{-1}{e}^{-i {\mathrm{e}}\theta}\,.
\end{equation}
There are, though,
many ways to construct such invariant fields that  are not
physically acceptable. In addition, then, we impose the kinematical
condition  that we want
charges that are, in fact, charged particles at early and late
times, i.e., they should have a sharp energy-momentum at these
asymptotic times. This condition is encoded in the dressing
equation:
\begin{equation}\label{dresseqtn}
h u^\mu\partial_\mu(h^{-1})=-i{\mathrm{e}}u^\mu A_\mu\,,
\end{equation}
where $u^\mu$ is the asymptotic four velocity of the particle.
This equation can be motivated either from the form of the
asymptotic dynamics associated with the dressed field, or from
heavy charge effective theory\,\cite{Horan}.

Equations (\ref{ginv}) and (\ref{dresseqtn}) are the fundamental
equations in our approach to charges
in QED and, as we will see later, QCD. For a given four velocity
$u^\mu$, which we write as $\gamma(\eta+v)^\mu$ where
$\eta=(1,0,0,0)$ and
$v=(0,\underline{v})$, we can solve these equations to obtain the
dressing
\begin{equation}
 h^{-1}=e^{-i{\mathrm{e}}K} e^{-i{\mathrm{e}}\chi}\,,
\end{equation}
where
\begin{eqnarray}
&&K(x)=-\int_{\Gamma}ds(\eta+v)^\mu\frac{\partial^\nu
F_{\nu\mu}}{{\mathcal G}\cdot\partial}\\ &&\chi(x)=\frac{{\mathcal
G}\cdot A}{{\mathcal G}\cdot\partial}
\end{eqnarray}
and ${\mathcal
G}^\mu=(\eta+v)^\mu(\eta-v)\cdot\partial-\partial^\mu$ while
$\Gamma$ is the past (future) trajectory of an incoming (outgoing)
particle. It is immediate from this form of the solution that the
dressing has structure:
\begin{itemize}
 \item a gauge dependent part associated with  $\chi$ which makes the  whole charge gauge
 invariant, we thus talk of this as the
 minimal part of the dressing;
\item a gauge invariant part associated with  $K$ which we call the  additional part of the
dressing. It is needed to satisfy the dressing equation.
\end{itemize}
To gain a better understanding of why these different structures
arise in the dressing and, indeed, to verify that the dressing equation
does capture the particle structure of the asymptotic charges as we expect, we
need to perform detailed perturbative tests of the construction.

In the present context, what we need to calculate are the on-shell
$n$-point Green's functions for the charged fields. Each charged
field needs to be dressed in the form appropriate to the
asymptotic velocity of its in-coming or out-going
particle. The dressing for each charge can then be expanded
perturbatively and the minimal and additional structure can be
seen to generate two new vertices. Armed with these new Feynman
rules,  the on-shell Green's functions can be calculated and we
find\,\cite{Bagan:1998kg} that, at the appropriate points on the
mass-shell:
\begin{itemize}
\item the infra-red divergences associated with the minimal part of the
dressing cancel those soft divergences that arise from the perturbative expansion
of the matter;
\item in pair creation processes, the  additional gauge invariant part of the
dressing generates a phase divergence which cancels the one which
arise from the matter fields.
\end{itemize}
Thus we have seen that, already at the level of Green's functions,
our construction  does describe charged particles and that the
structure in the dressing reflects the differing infra-red
behaviour of the matter fields. In this context, we can view the
minimal part of the dressing as the soft component of the charge, and the
additional part as the phase.

\section*{Charges in QCD}

In contrast to the situation in QED where the electric charge was
itself gauge invariant, in an unbroken non-abelian gauge theory,
such as QCD, the very definition of charge requires a restriction
on the form of the gauge transformations at spatial
infinity\,\cite{Lavelle:1996tz,Lavelle:1997ty}. Once this is
correctly taken into account, though, the route to the
construction of non-abelian charged particles proceeds in a now
familiar manner:
\begin{itemize}
\item {gauge invariance is maintained through the dressing
$h^{-1}$ where now, under a non-abelian gauge transformation,
$h^{-1}(x)\to h^{-1}(x)U(x)$;}
  \item {the kinematical restriction required to describe an asymptotic charged particle with
  four velocity $u$ is
  $h u\cdot\partial(h^{-1})=gu\cdot A$.}
\end{itemize}
All that needs to be done, then, is to solve these equations!

We have presented an algorithm for the perturbative construction
of the soft part of the dressing to an arbitrary order in
perturbation theory and we can  incorporate some non-perturbative
effects\,\cite{Lavelle:1997ty}. We have also proven  that there is
a global obstruction to the construction of such dressings. This
follows because it can be shown that any such dressing could be
used to construct a gauge fixing which, however,  we know cannot
exist globally. This shows that isolated quarks are not
observables and colour must be confined.

In addition to the construction of charged matter, in a
non-abelian theory we also need to construct gauge invariant glue.
The dressing can also be used for this and by analyzing what
happens in massless QED we can see that the dressing equation
  for a massless charge will also lead to a vanishing  asymptotic
Hamiltonian\,\cite{Horan}.  This gives us confidence that even the
collinear structure associated with such charges will be correctly
captured in our approach.

Rather than discuss these general results here, it is perhaps more helpful to present a concrete
calculation which captures the essential new ingredients found in the non-abelian
theory. This will also allow us to give a new physical
interpretation to the minimal part of the dressing. Thus we shall conclude this talk
by presenting a direct route to the calculation of the  potential between static
quarks.

\section*{The Interquark Potential}

Up to now we have essentially been concerned with the construction
of isolated charges. However, when it comes to hadrons we are
faced with the question of whether these colour singlets have a
constituent structure, i.e.,  are they  constructed from
individual gauge invariant constituent charges? There are various
models for hadronic structure, some of which have a constituent
structure and some of which do not. We wish to investigate to what
extent these pictures follow from QCD.

The method we are going to follow is to construct gauge invariant
constituent states and then study their energy. This will then be
compared with that derived from the Wilson loop which we
recall\,\cite{Fischler:1977yf,Appelquist:1977tw} yields, at order
$\alpha^2$, the following potential between two static quarks
\begin{equation}
V(r)= -\frac{g^2 C_F}{4\pi r}\left[ 1+
\frac{g^2}{4\pi}\frac{C_A}{2\pi }\left( {4}-{\frac13} \right)
\log(\mu r) \right]\,.
\end{equation}
At lowest order this is a Coulombic potential and the logarithmic
correction we have split here into two parts reflecting the
anti-screening and screening contributions typical of a
non-abelian gauge theory.

The dominant anti-screening contribution comes from longitudinal
glue and the screening part from gauge invariant
glue\,\cite{Lavelle:1998dv}. This divide has been seen in a wide
variety of studies of the $\beta$-function. Since our dressings
factorise into two parts (a minimal part constructed out of
longitudinal degrees of freedom and an additional gauge invariant
term) we would expect that if a constituent picture holds then the
minimal part of the dressing for the constituents should reproduce
the dominant anti-screening term in the potential. We will now
show this.

To calculate the potential at order $\alpha^2$ we will need to
extend the  minimal dressing to $O(g^3)$. Having done this the
potential is the separation dependent part of the expectation
value of the Hamiltonian between the   charged constituent states.

The minimal static dressing in QED was: $\exp(-ie\chi)$, with
$\chi={\partial_i A_i}/{\nabla^2}$. In QCD this has the gauge
invariant extension
\begin{equation}
\exp(-ie\chi)\Rightarrow \exp(g\chi^aT^a)\equiv h^{-1}
\end{equation}
with $g\chi^aT^a=(g\chi_1^a+g^2 \chi_2^a+g^3\chi_3^a+\cdots)T^a$
where
\begin{equation}
\chi_1^a=\frac{\partial_j A_j^a}{\nabla^2}\,;\quad
\chi_2^a=f^{abc}\frac{\partial_j}{\nabla^2}\left(
\chi_1^bA_j^c+\frac12(\partial_j\chi_1^b)\chi^c_1
\right)\,;\quad\dots
\end{equation}
Note that for the rest of this talk we denote by $h^{-1}$ the
above minimal static dressing.

Now we  take the expectation value of the Hamiltonian between
such minimally dressed quark/antiquark states:
$\bar\psi(y)h(y)h^{-1}(y')\psi(y')\ket0$. This yields for the
potential
 \begin{equation}
  -\textrm{tr}\int\!d^3x \,\bra0 [E^a_i(x),h^{-1}(y)]h(y)
[E^a_i(x),h^{-1}(y')]h(y') \ket0
\end{equation}
At $O(g^2)$ we obtain for $ V^{g^2}(r)$
\begin{equation}
  -g^2\textrm{tr}\!\int\!d^3x \bra0 [E^a_i(x),
  \chi^d_1(y)]T^dT^b
  [E^a_i(x),\chi^b_1(y')] \ket0\,.
 \end{equation}
From the equal-time commutator:
$[E_i^a(x),A_j^b(y)]=i\delta^{ab}\delta_{ij}
\delta(\boldsymbol{x}-\boldsymbol{y})$ we thus obtain the expected
Coulomb potential:
\begin{equation}
  V^{g^2}(r)=
  -\frac{g^2 N C_F}{4\pi r}\,.
\end{equation}

At $O(g^4)$ we  require $[E^a_i(x),h^{-1}(y)]h(y)$. From simple
manipulations we obtain
\begin{equation}
 g[E_i^a(x),\chi_1(y)]
 +g^2[E_i^a(x),\chi_2(y)]+g^3 [E_i^a(x),\chi_3(y)]
 +O(g^4)\,,
\end{equation}
\noindent plus many gauge dependent terms which we can drop. This
yields for the potential\,\cite{Lavelle:1998dv}
\begin{eqnarray}
 V^{g^4}(r)&\!=&
\! - \frac{3g^4 C_FC_A}{(4\pi)^3}\int\!d^3z\!\int\!d^3w
 \frac1{\mod{\boldsymbol z-\boldsymbol w}} \times \nonumber\\
&& \qquad\left(  \partial_j^z\frac1{\mod{\boldsymbol z-\boldsymbol
y}}\right)
 \left(  \partial_k^w\frac1{\mod{\boldsymbol w-\boldsymbol {y}'}}\right)
\langle A_k^T(w)A_j^T(z)\rangle\,.
\end{eqnarray}
Inserting the free propagator
\begin{equation}
\langle A_k^T(w)A_j^T(z)\rangle=
\frac1{2\pi^2}\frac{(z-w)_j
(z-w)_k}{{\mod{\boldsymbol z-\boldsymbol w}}^4}
\end{equation}
and performing the integrals, we easily find
\begin{equation}
V^{g^4}(r)=-\frac{g^4}{(4\pi)^2}\frac{NC_FC_A}{2\pi
r}\,{4}\log(\mu r)\,.
\end{equation}
This we recognise as the expected  anti-screening contribution to
the interquark potential. This calculation has shown that the
non-abelian anti-screening effect can be understood as arising
from the interaction of two gauge invariant constituents.

\section*{Conclusions}

Starting from the observation that in gauge theories the
asymptotic fields are not free, we have seen that Gauss' law
implies that dressings are needed to construct physical charged
fields. We have presented a method of constructing dressed charges
and have seen that this results in a structured dressing.

In QED we have demonstrated that the dressings remove the
infra-red  divergences associated with on-shell Green's functions.
We note that the  different structures found in the  dressings
play different roles in this cancellation.

In non-abelian gauge theories such as QCD well-defined colour
charges must be dressed if colour is to be a good quantum number.
The perturbative construction of dressed fields has been
investigated by various authors\,\cite{haller:1997}, however, we
stress that there is a global obstruction to the construction of
an isolated colour charge.

To test the physical validity of constituent structures we have
studied the potential between such gauge invariant quarks. The
anti-screening contribution to the full interquark potential was
recovered by considering minimally  dressed quarks. This result
shows that we have determined the dominant glue configuration and
renders a constituent structure visible.

Extensions of this work to higher orders in the potential and to
the usual screening contribution are in progress. Given the global
obstruction to the construction of dressings, the above
constituent picture of hadrons must  break down at some scale. The
major challenge of this  programme of research is to determine
this scale.

\section*{Acknowledgements} This work was supported in part by the
British Council and the  Acciones Integradas (HB 1997-0141)
programme.

\section*{References}

%\bibliographystyle{h-physrev}
%\bibliography{litbank1}

\begin{thebibliography}{1}

\bibitem{kulish:1970}
P.~Kulish and L.~Faddeev,
\newblock Theor. Math. Phys. {\bf 4}, 745 (1970).

\bibitem{Horan}
R.~Horan, M.~Lavelle, and D.~McMullan{, \textit{Charges in gauge
theories}},
\newblock (1998),
\newblock Plymouth preprint PLY-MS-48.

\bibitem{Bagan:1998kg}
E.~Bagan, M.~Lavelle, and D.~McMullan,
\newblock Phys. Rev. {\bf D57}, 4521 (1998), hep-th/9712080.

\bibitem{Lavelle:1996tz}
M.~Lavelle and D.~McMullan,
\newblock Phys. Lett. {\bf B371}, 83 (1996), hep-ph/9509343.

\bibitem{Lavelle:1997ty}
M.~Lavelle and D.~McMullan,
\newblock Phys. Rept. {\bf 279}, 1 (1997), hep-ph/9509344.

\bibitem{Fischler:1977yf}
W.~Fischler,
\newblock Nucl. Phys. {\bf B129}, 157 (1977).

\bibitem{Appelquist:1977tw}
T.~Appelquist, M.~Dine, and I.~J. Muzinich,
\newblock Phys. Lett. {\bf 69B}, 231 (1977).

\bibitem{Lavelle:1998dv}
M.~Lavelle and D.~McMullan,
\newblock (1998), hep-th/9805013,
\newblock to appear in Phys. Lett. B.

\bibitem{haller:1997}
L.~Chen, M.~Belloni, and K.~Haller,
\newblock Phys. Rev. {\bf D55}, 2347 (1997), hep-ph/9609507,
\newblock see also talk by K. Haller at this conference.

\end{thebibliography}

\end{document}